\documentstyle[eqsecnum,aps]{revtex}
\begin{document}
%
%
%
%
%
%\mbox{ }\hfill{\normalsize UCT-TP 236/97}\\
%\mbox{ }\hfill{\normalsize \today}\\
%
%
%
%
%
\title{ Exact Baryon, Strangeness and 
Charge Conservation \\[.5cm]
in Hadronic Gas Models.} 
\author{J. Cleymans,  M. Marais}
\address{Department of Physics, University of Cape Town, Rondebosch 7700,\
South Africa}
\author{E. Suhonen}
\address{Department of  Physical Sciences,
University of Oulu, FIN-90571 Oulu, Finland.}
\maketitle
\begin{abstract}
Relativistic  heavy  ion  collisions  are studied assuming that
particles  can  be  described  by  a  hadron gas in thermal and
chemical  equilibrium. The exact conservation of baryon number,
strangeness and charge are explicitly taken into account. For
heavy  ions  the  effect  arising  from the neutron surplus
becomes
important and  leads to a
substantial increase in e.g. the $\pi^-/\pi^+$ ratio. 
A method is developed which is very well suited for the study 
of small systems.
\end{abstract}
 \section{Introduction}
It  has  been  shown  recently in a very convincing manner that
hadronic   matter  produced  in  collisions  of  small  systems
($e^+e^-$, $p-p$, $p-\overline{p}$) is very close to 
chemical  equilibrium. 
 With  only  a  small  number  of
parameters, namely the temperature $T$, the volume $V$ and, a
factor $\gamma_S$ which measures the deviation of strange particles
from chemical equilibrium, it is possible to fit close to 30
different particle abundances~\cite{becattini1,becattini2}. 
It  is  of great interest to establish whether or not this also
holds for  systems of medium size like $S-S$ and for
relativistic  heavy  ion collisions like $Au-Au$ at the BNL-AGS
or  $Pb-Pb$  at CERN~\cite{qm96}. 
This is of relevance because it is widely
believed   that   in   these  collisions  one  could  create  a
quark-gluon  plasma.  Of  immediate  interest  is  the observed
increase  in the abundance of strange particles
in heavy ion collisions. In this 
paper we would like to present a method for 
taking into account the exact conservation of
baryon number $B$, strangeness $S$ and charge $Q$. 
The  method  is  ideally  suited  to analyze the extension from
small
systems like  $p-p$ towards bigger systems up to $S-S$.  For
larger systems 
the method becomes impractical 
and should be replaced by a more straightforward approach based
on the grand canonical ensemble.
Our numerical results can be compared with data from 
the Brookhaven National Laboratory, 
e.g. the E866 collaboration~\cite{E866} which
measures 
the dependence of  hadronic ratios on the number
of  projectile  participants
thereby  allowing one to  study the transition from small to large
systems in a systematic way.
These results  
give insight into the behavior of the produced hadronic
system as a function  
its size.
The  
treatment 
presented in this paper differs 
from 
the standard one based on the grand canonical ensemble
(see for example~\cite{stachel1,stachel2,jc-de} 
in that we consider the 
quantum number  content exactly.  This means that we do not 
introduce chemical potentials for
the baryon number or for strangeness. Chemical potentials are usually 
introduced to enforce the right quantum numbers of the system in
an average sense. This is a correct treatment for large systems, however,
for  small systems the production of   extra  
proton - anti-proton pairs  for example
will clearly be more suppressed than in  large systems. 
These extra corrections
were first pointed out by
Hagedorn~\cite{hagedorn}  and subsequently a complete treatment
has been 
formulated
\cite{turku,hagedorn-redlich,muller,greiner,gorenstein,heinz,cley1,cley3}.
We emphasize that these corrections 
do  not contain information about the dynamics. They simply
follow from baryon number conservation and 
 must be taken into account before
considering  more involved models. It is also worth emphasizing
that they
do not introduce any new parameters.
Our treatment differs from the one presented in~\cite{becattini1,becattini2}
in that it is much more analytic. It extends the analysis started 
in~\cite{jc-atm} since it 
also includes the exact treatment of charge conservation 
and allows for 
the correct treatment of systems which are not isospin-symmetric.
\section{Partition Function}
The exact treatment of quantum numbers in statistical mechanics 
has been well  
established for some time~\cite{turku}. It is obtained by projecting
the partition function onto the desired values of $B$, $Q$ and  $S$
\begin{equation}
Z_{B,Q,S}={1\over 2\pi}\int_0^{2\pi}d\phi\ e^{-iB\phi}
\;  {1\over 2\pi}\int_0^{2\pi}d\alpha\ e^{-iQ\alpha}
{1\over 2\pi}\int_0^{2\pi} d\psi  e^{-iS\psi}
Z(T,\lambda_B,\lambda_Q,\lambda_S)
\end{equation}
where 
$Z$ is the (grand canonical) partition function
and the usual fugacity 
factors $\lambda_B$ and $\lambda_S$ have been replaced by :
\begin{equation}
\lambda_B = e^{i\phi}~~~~~ \lambda_Q = e^{i\alpha}~~~~~~ \lambda_S = e^{i\psi}.
\end{equation}
As  the  contributions  always  come  pairwise for particle and
anti-particle,
the fugacity factors will give rise to the cosine of the angle.
In the  extended treatment it is useful to
group all particles appearing in the Particle  Data Booklet~\cite{PDB}
into fourteen categories
depending on their quantum numbers (we leave out charm and bottom).
$Z_K$ is the sum (given below) of 
all mesons having strangeness $\pm 1$ ($K,\bar{K}, K^*,\dots $)
and zero charge. Similarly
$Z_N$ is the sum of all baryons and anti-baryons having zero strangeness,
$Z_Y$ is the sum of all hyperons and anti-hyperons,
 while $Z_0$ is the sum of all
non-strange mesons.
%We do not include cascade particles as their 
%contribution is unimportant for the energy range 
%under consideration and their inclusion considerably
% complicates the formalism.\\
%
\vskip 0.5 truecm
\begin{center}
\begin{tabular}{|c|c|c|}
\hline  
Quantum Numbers       &  Lowest Mass Particle & Notation     \\ 
\hline  
$S=0~ B=0~ Q=0 $   &  $\pi^0$            & $Z_0 $          \\
$S=0~ B=0~ Q=1 $   &  $\pi^+$            & $Z_{\pi^c}$     \\
$S=0~ B=1~ Q=-1$   &  $\Delta^- $        & $Z_{\Delta^-}$   \\
$S=0~ B=1~ Q=0 $   &  $n$                & $Z_n$           \\
$S=0~ B=1~ Q=1 $   &  $p$                & $Z_p$           \\
$S=0~ B=1~ Q=2 $   &  $\Delta^{++}$      & $Z_{\Delta^{++}}$\\
$S=1~ B=0~ Q=0 $   &  $K^0$   & $Z_{K}$         \\
$S=1~ B=0~ Q=1 $   &  $K^+$              & $Z_{K^c}$       \\
$S=-1~ B=1~ Q=1 $  &  $\Sigma^+ $& $Z_{\Sigma^+}$  \\
$S=-1~ B=1~ Q=-1 $ &  $\Sigma^-$& $Z_{\Sigma^-}$  \\
$S=-1~ B=1~ Q=0 $  &  $\Lambda   $       & $Z_{\Lambda}$   \\
$S=-2~ B=1~ Q=0 $  &  $\Xi^0     $       & $Z_{\Xi^0}$   \\
$S=-2~ B=1~ Q=-1 $  &  $\Xi^-    $        & $Z_{\Xi^-}$   \\
$S=-3~ B=1~ Q=-1 $  &  $\Omega   $        & $Z_{\Omega}$   \\
\hline 
\end{tabular}
\end{center}
\vskip 0.5 truecm
\begin{center}
{\it Table 1 : Classification of the families of particles.}
\end{center}
\vskip 0.5 truecm
For example, 
$Z_{K^c}$ is the single particle partition function of all charged 
strange mesons ($K^+, K^-, K^{*+}, K^{*-}$...) i.e.
\begin{equation}
Z_{K^c}= \sum_{j\in |S|=1, |B|=0, |Q|=1} g_j V\int {d^3p\over (2\pi)^3}
e^{-E_j/T} ,
\end{equation}
where $g_j$ is the degeneracy factor.

With the help of the definitions presented in table 1,
the partition function can be rewritten
in the following form,
\begin{eqnarray}
Z_{B,S,Q}\left( T,V\right) &=& {1\over (2\pi)^3}
\int_0^{2\pi }d\phi\; \exp \left( -iS\phi \right) \int_0^{2\pi }d\psi\; \exp
\left( -iB\psi \right) \int_0^{2\pi }d\alpha\; \exp \left( -iQ\alpha \right)
\nonumber \\
&&\cdot \exp \left[ 2Z_K\cos \phi +2Z_n\cos \psi +2Z_{\pi^c}\right.
\cos \alpha +2Z_{\Lambda}\cos(\psi -\phi) 
\nonumber\\
&&+2Z_{K^c}\cos \left( \phi +\alpha \right) 
  +2Z_{\Delta^-}\cos \left( \psi -\alpha \right) 
  +2Z_p\cos \left(\psi +\alpha \right) 
  +2Z_{\Delta^{++}}\cos \left( \psi +2\alpha \right) 
\nonumber \\
&&+2Z_{\Sigma^{+}}\cos \left( -\phi +\psi +\alpha \right) 
  +2Z_{\Sigma^{-}}\cos \left( -\phi +\psi -\alpha \right) 
\nonumber\\
&&+2Z_{\Xi^-}\cos \left( -2\phi +\psi -\alpha \right) 
  +2Z_{\Xi^0}\cos \left( -2\phi +\psi \right) 
\nonumber\\
&&+\left. 2Z_{\Omega}\cos \left( -3\phi +\psi -\alpha \right) \right]
\cdot \exp (Z_0)   .
 \label{bsq2}
\end{eqnarray}
To  continue  we  would  like  to  make  use  of  the  integral
representation of the modified Bessel function
\begin{equation}
I_n(z) ={1\over \pi}\int_0^{\pi} e^{z\cos\theta}\cos{n\theta}  .
d\theta
\end{equation}
However,  this cannot be used in a straightforward manner since
the  dependence  on  angles  does  not factorize and almost all
terms   involve   more  than  one  angle.  To  circumvent  this
difficulty  we  introduce  new  angles  whenever  more than one
appears.  For  example,  for  the  term involving $Z_\Lambda$ we
introduce an intermediate angle $\lambda$ in the following way 
\begin{eqnarray}
1&=&\int_0^{2\pi}
d\lambda~\delta(\psi-\phi-\lambda)\nonumber\\
 &=&\sum_{n=-\infty}^\infty {1\over 2\pi}\int_0^\pi 
          d\lambda~e^{in(\psi-\phi-\lambda)}
\end{eqnarray}
With  this method complete factorization can be achieved at the
price  of introducing a new summation at each step. Since there
are ten cosines in Eq.~(2.4) involving more than one angle, we end
up    having   ten   summations.   The   
following result   is  
obtained~\footnote{Full details can be found in ~\cite{marais}.}
\begin{eqnarray}
Z_{B,S,Q}\left( T,V\right) 
&=&\left[\prod_{j=1}^{10} \sum_{n_j=-\infty }^\infty\right] 
I_{\nu_S}\left( 2Z^{K^0}\right)I_{\nu_B}\left( 2Z^n\right) 
I_{\nu_Q}\left( 2Z_{\pi^c}\right) \nonumber \\
&&I_{-n_1 }\left( 2Z_{\Lambda ^0}\right)
I_{-n_2}\left( 2Z_{K^c}\right)  I_{-n_3}\left( 2Z_p\right) \nonumber \\
&&I_{-n_4}\left( 2Z_{\Delta^{-}}\right)
I_{-n_5}   \left( 2Z_{\Sigma^{+}}\right) 
I_{-n_6}   \left( 2Z_{\Sigma^{-}}\right) 
I_{-n_7}   \left( 2Z_{\Xi ^0}    \right)  \nonumber \\
&&I_{-n_8} \left( 2Z_{\Xi^{-}}   \right) I_{-n_9}\left( 2Z_\Omega \right)
I_{-n_{10}}\left(2Z_{\Delta^{++}}\right) \cdot \exp \left( Z_0\right)
 \label{bsq17}
\end{eqnarray}
where
\begin{eqnarray}
\nu_S&=&S-n_1 +n_2-n_5-n_6-2n_7-2n_8-3n_9        \\
\nu_B&=&B+n_1 +n_3+n_4+n_5+n_6+n_7+n_8+n_9+n_{10}\\
\nu_Q&=&Q+n_2+n_3-n_4+n_5-n_6-n_8-n_9+2n_{10}            .
\end{eqnarray}
Equation (2.7) is our main result and forms the basis 
of our analysis below. The sums in the
above  equation  are  dominated by 
terms corresponding to the index  of the Bessel
functions 
being zero (for the size of the system being sufficiently small). This
means  however  that  for large values of the baryon number $B$
some  of  the indices could become quite large. In
such  cases  evaluation  of  the sums becomes very
time-consuming  numerically  and  it is better to resort to the
grand canonical ensemble. 

The canonical partition function for a gas with three conserved quantum
numbers has thus been derived and will be used in the following to
derive expressions for particle numbers.
The differentiation of the equation (2.7) for particle 
abundances decreases/increases some of the 
indices $\nu_S, \nu_B$ and $\nu_Q$ by one and 
leads to useful R-factors which, for different particle 
species deviate from
$Z_{B,S,Q}(T,V)$ in the  way indicated below
\begin{equation}
\begin{array}{llll}
R_{K^+/K^-}: && \nu_S \rightarrow \nu_S\mp 1; &\nu_Q \rightarrow \nu_Q\mp 1 \\
R_{K^0/\overline{K^0}}: && \nu_S \rightarrow \nu_S\mp 1            \\
R_{p/\bar{p}}: & \nu_B \rightarrow \nu_B\mp 1; && \nu_Q \rightarrow \nu_Q\mp 1
\\ 
R_{\Lambda/\bar{\Lambda}}: & \nu_B \rightarrow \nu_B\mp 1; &
	 \nu_S \rightarrow \nu_S\pm 1& \\
R_{\Sigma^+/\bar{\Sigma^+}}:  & \nu_B \rightarrow \nu_B\mp 1;
	& \nu_S \rightarrow \nu_S\mp 1;& \nu_Q \rightarrow \nu_Q\mp 1 \\
\end{array}
\end{equation}

If a particle, $i$, has strangeness 1, baryon number 0 and charge 0,
 it's density will be given by
\begin{equation}
n_{i} = \left[{R_K\over Z_{B,S,Q}}\right]
g_i \int {d^3p\over (2\pi)^3}e^{-E_i/T} ,
\end{equation}
with 
\begin{eqnarray}
R_K 
&=&\left[\prod_{j=1}^{10} \sum_{n_j=-\infty }^\infty \right]
I_{\nu_S -1}\left( 2Z^{K^0}\right)I_{\nu_B}\left( 2Z^n\right) 
I_{\nu_Q}\left( 2Z_{\pi^c}\right) \nonumber \\
&&I_{-n_1 }\left( 2Z_{\Lambda ^0}\right)
I_{-n_2}\left( 2Z_{K^c}\right)  I_{-n_3}\left( 2Z_p\right) \nonumber \\
&&I_{-n_4}\left( 2Z_{\Delta^{-}}\right)
I_{-n_5}   \left( 2Z_{\Sigma^{+}}\right) 
I_{-n_6}   \left( 2Z_{\Sigma^{-}}\right) 
I_{-n_7}   \left( 2Z_{\Xi ^0}    \right)  \nonumber \\
&&I_{-n_8} \left( 2Z_{\Xi^{-}}   \right) I_{-n_9}\left( 2Z_\Omega \right)
I_{-n_{10}}\left(2Z_{\Delta^{++}}\right) \cdot \exp \left( Z_0\right) .
 \label{bsq18}
\end{eqnarray}
All other particle densities are obtained by using the appropriate $R$ factor.
The factor in square brackets in equations (2.12) and (2.13) replaces 
the fugacity  in the usual
grand canonical ensemble treatment~\cite{stachel1,stachel2,jc-de}.
Having thus determined all particle 
densities, we consider
the behavior at freeze-out time. In this case all the resonances 
in the gas are allowed to decay 
into lighter stable particles. 
This means that each particle density is multiplied with
its appropriate branching ratio (indicated by $Br$ below). 
The abundances of  particles in the final state are thus
determined by :
\begin{equation}
n_H = \sum_i n_i Br(i\rightarrow H)  
\end{equation}
where  each  sum  runs  over  all  particles  contained  in the
hadronic gas and $H$ refers to a hadron ($\pi^+, K^+, \dots$).
\section{Numerical Results}
In  order to stay close to the grand canonical ensemble 
treatment we keep
the  temperature  and the density $B/V$ fixed.
The latter corresponds
to  keeping  the  baryon chemical potential fixed. We therefore
expect  all ratios to become constant as the size of the system
is increased.
Fig. 1 illustrates the approach to 
the  thermodynamic (grand canonical) limit for 
the $K^+/\pi^+$ ratio. As the system
becomes  larger  the  ratio  approaches a constant. 
The variation of  
 the  $\pi^-/\pi^+$  ratio as one increases the neutron surplus
 or, equivalently, the ratio $B/2Q$, is shown in Fig. 2.
 As expected one observes a  clear increase of this ratio with
$B/2Q$. 
In  Fig.  3  we  show  the $p/\pi^+$ ratio as a function of the
baryon number $B$.
The $K^+/K^-$ ratio is shown in Fig. 4 for a fixed value of the
temperature,  $T=  110$ MeV. One observes a very smooth decrease
and this ratio 
very quickly approaches its asymptotic value.

Fig.  5  shows  the  $\bar{p}/\pi^+$ ratio as a function of the
size  of  the  system.  This  figure  is interesting because 
preliminary  results  from  the  E866 collaboration~\cite{E866}
show  a  {\it  decrease}  of  this ratio with increasing system
size,  i.e.  the opposite  behavior from the one 
observed in Fig. 5.
 In our opinion, this proves in a convincing manner that $\bar{p}$ are
definitely   not   in    thermal
equilibrium. It would be difficult to show this if the analysis
is based on only one system.
\section{Summary}
We have assumed that, as  a first approximation, the hadronic final 
state produced in a relativistic heavy ion collision can be described 
by a hadronic gas which is in chemical equilibrium. 
The standard description of such a system makes use of 
a statistical description based on the grand canonical ensemble
using chemical potentials for the conserved quantum numbers.
However,
if the system is very small then there will 
{\it always} be corrections due to the size of the system.
We have presented a method for taking into account 
corrections arising from the exact conservation of
baryon number $B$, strangeness $S$ and, charge $Q$. This method is 
well suited for 
small systems and numerical results have been presented
for the most abundantly observed hadrons. 
For large systems the 
numerical evaluation  becomes 
very slow 
because a large number of terms have to be kept in the sums.
The numerical results presented can be compared with data from 
Brookhaven,  e.g.  the  E866 collaboration~\cite{E866}. We 
plan to do this in the near future.
\\
\subsection*{Acknowledgment}
\hspace*{\parindent}
We grate\-ful\-ly acknowledge  
useful and stimulating
discussions with Helmut Satz, 
Krzysz\-tof Red\-lich 
and Di\-nesh Sri\-vastava.

\begin{center}
FIG. 1. The $K^+/\pi^+$
ratio as a function of 
the baryon number $B$ for  fixed  values  of  the
temperature $T$, the baryon density $B/V$ and $B/2Q$.
\end{center}
\begin{center}
FIG. 2. The $\pi^-/\pi^+$ ratio as a function of the
$B/2Q$ 
for  a fixed  value  of  the
temperature  $T$,  the  baryon  density  $B/V$,  and the baryon
number ($B=10$).
\end{center}
\begin{center}
FIG. 3. The $p/\pi^+$
ratio as a function of 
the baryon number $B$ for  fixed  values  of  the
temperature $T$, the baryon density $B/V$ and $B/2Q$.
\end{center}
\begin{center}
FIG. 4. The $K^+/K^-$ ratio as a function of the
the baryon number $B$ for  fixed  values  of  the
temperature $T$, the baryon density $B/V$ and $B/2Q$.
\end{center}
\begin{center}
FIG. 5. The $\bar{p}/\pi^+$
ratio as a function of 
the baryon number $B$ for  fixed  values  of  the
temperature $T$, the baryon density $B/V$ and $B/2Q$.
\end{center}
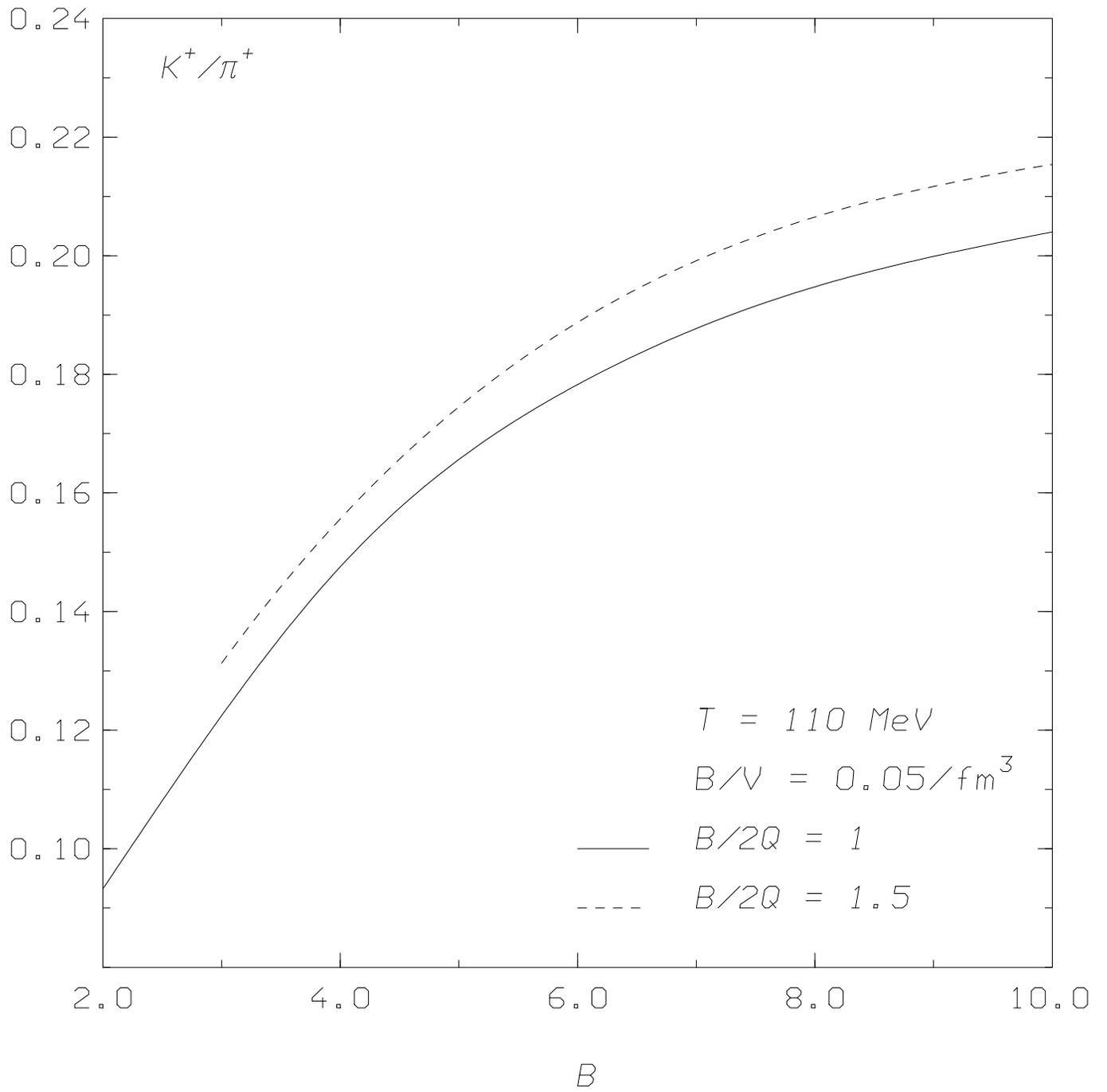
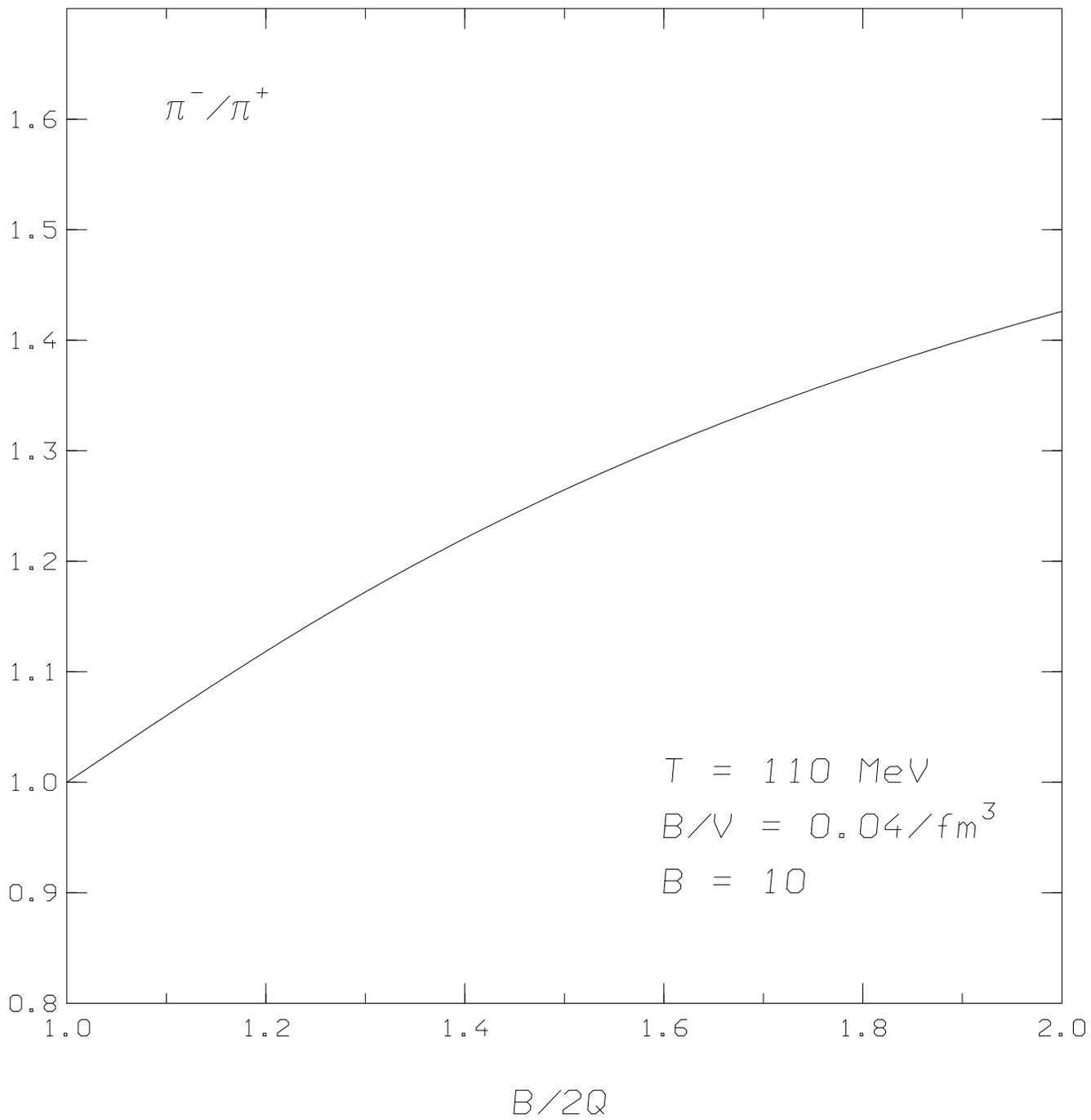
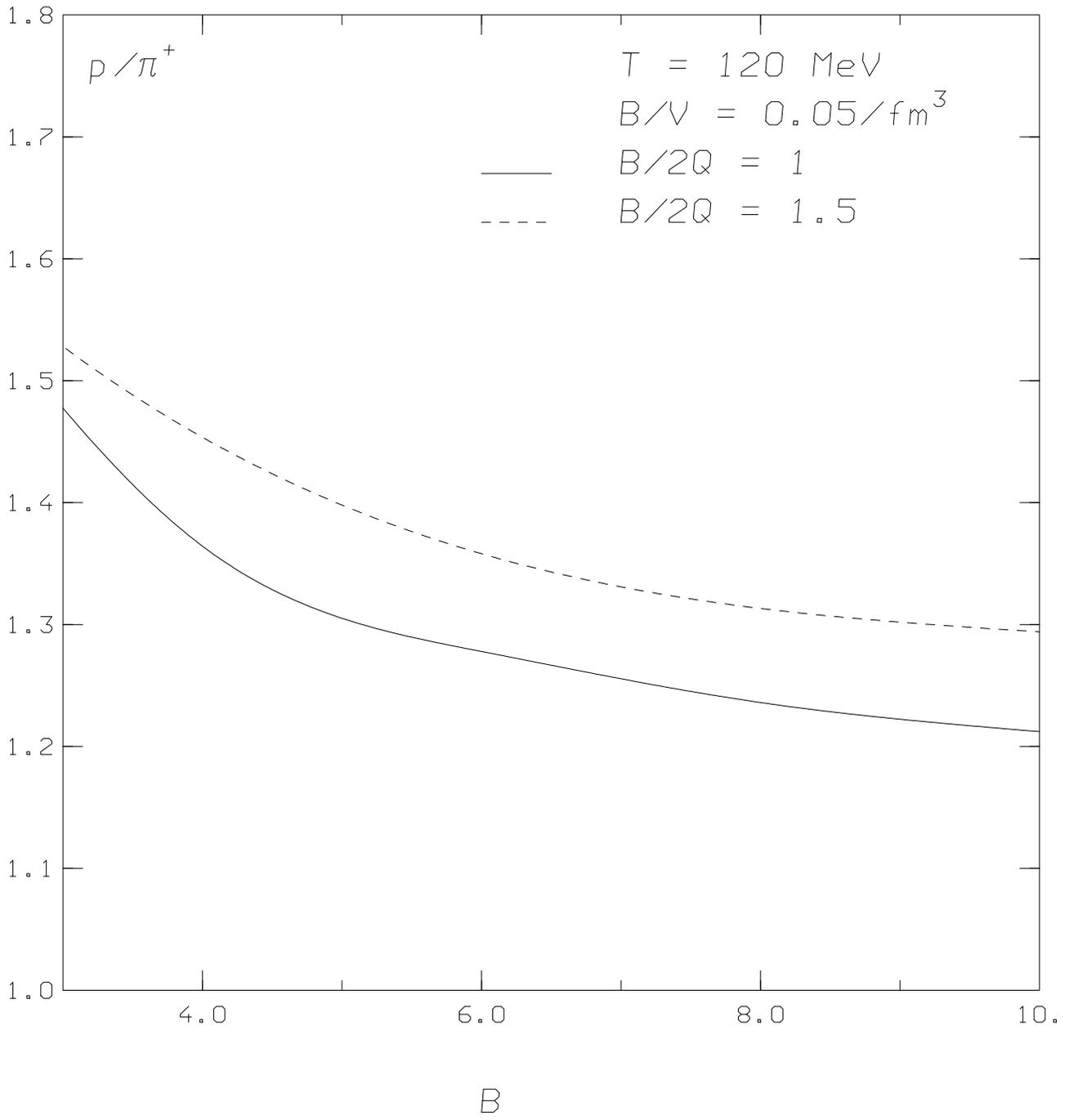
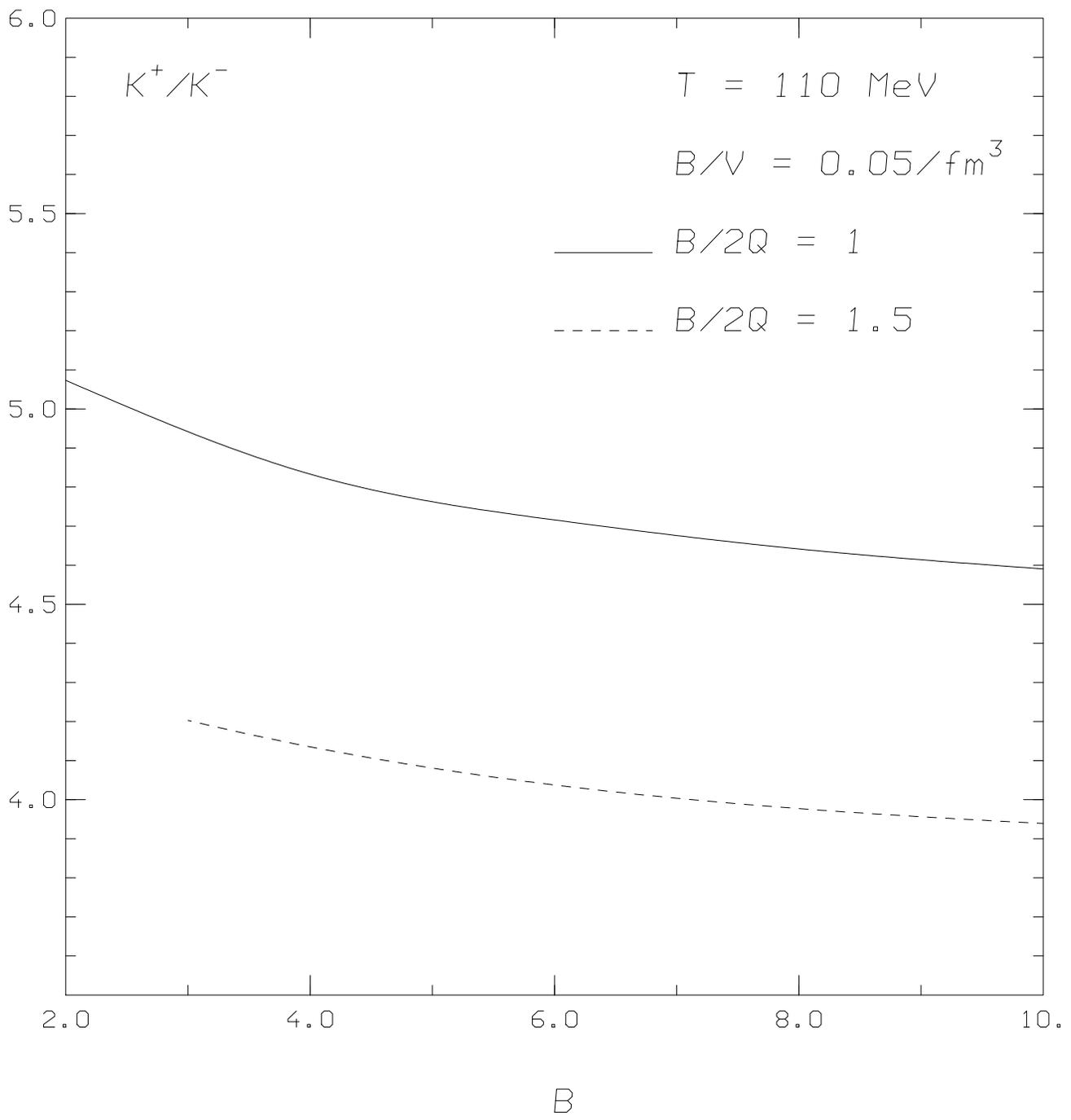
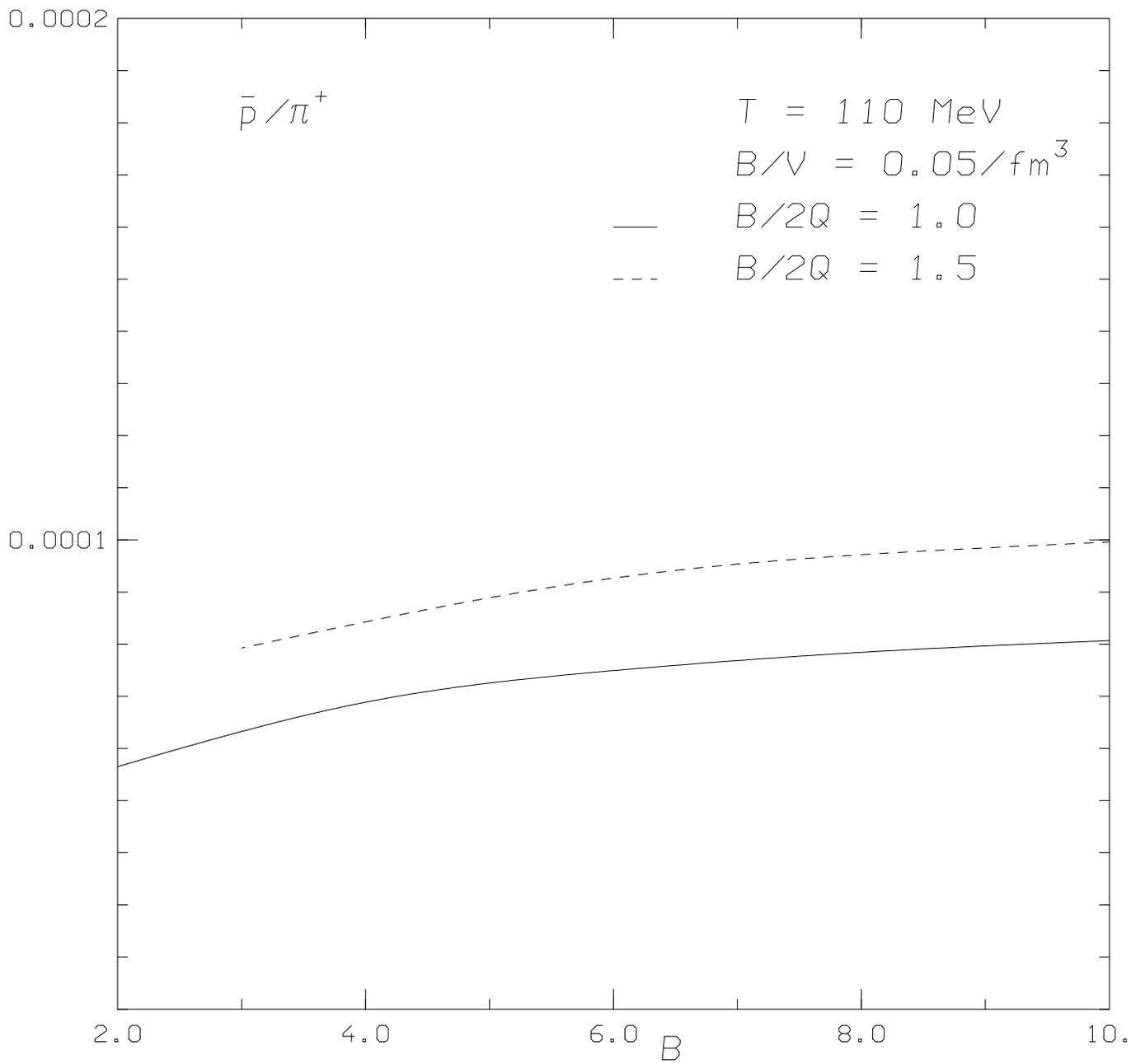
\end{document}